\documentclass[proceedings]{JHEP3}

\PrHEP{PrHEP unesp2002}
\conference{Workshop on Integrable Theories, Solitons and Duality}
\usepackage{epsfig, amsmath, verbatim}

\title{Quantum group symmetry of integrable models on the half-line}

\author{\speaker{Gustav W Delius} and Alan George\\
        Department of Mathematics,
        The University of York, York YO10 5DD,UK\\
        E-mail: \email{gwd2@york.ac.uk} and \email{ag160@york.ac.uk}}

\abstract{
    We derive the non-local conserved charges in the sine-Gordon
    model and affine Toda field theories on the half-line. They
    generate new kinds of symmetry algebras that are coideals of
    the usual quantum groups. We show how intertwiners of tensor product
    representations of these
    algebras lead to solutions of the reflection equation.
    We describe how this method for finding solutions to the reflection
    equation parallels the previously known method of using
    intertwiners of quantum groups to find solutions to the
    Yang-Baxter equation. }

\begin{document}

\section{Introduction}
\label{Intro}

During the last year the search for the quantum group symmetry of
integrable quantum field theories with boundaries has finally
succeeded\footnote{Earlier calculations for the sine-Gordon model
at the free fermion point were done in \cite{Mezincescu}}. MacKay
and Short in their contribution to these proceedings \cite{MacKay}
discuss the non-local symmetry charges of the principal chiral
model on the half-line. In this contribution we present the
non-local charges of the sine-Gordon model and affine Toda field
theories on the half-line.

These non-local charges generate symmetry algebras of a very
unusual kind. In contrast to all other known symmetry algebras
they are not bialgebras. We will refer to them as boundary quantum
groups. They are coideals of the usual quantum groups.  Just as
quantum groups lead to solutions of the Yang-Baxter equation, our
new boundary quantum groups lead to solutions of the reflection
equation, also known as the boundary Yang-Baxter equation.

The Yang-Baxter equation appears throughout 1+1 dimensional
integrable systems in two roles. Firstly to define commuting
transfer matrices one requires L operators that satisfy the
Yang-Baxter equation \cite{Baxter}. Secondly particles scatter in
a factorizable way, the consistency condition on the factorization
is given by the Yang-Baxter equation \cite{Zamolodchikov2}. In
section~\ref{YBE} we will review the Yang-Baxter equation in the
specific context of particle scattering and we will recall how
quantum group symmetry provides an easy way to find its solutions
\cite{Jimbo}.

In order to restrict the domain of an integrable system to the
half line or an interval one must use solutions of the reflection
equation in addition to solutions of the Yang-Baxter equation.
Such reflection matrices are used to describe factorizable
scattering of particles from the boundary \cite{Cherednik,
Ghoshal}. Reflection matrices are also needed for the definition
of commuting transfer matrices \cite{Sklyanin}. We describe the
reflection equation in section~\ref{bYBE}, emphasizing how it
parallels the Yang-Baxter equation. We show how boundary quantum
groups allow us to find solutions of the reflection equation by
simply solving a linear intertwining equation.

In section~\ref{sG} we briefly review the non-local charges of the
sine-Gordon model \cite{Bernard} that generate the affine algebra
$U_q(\widehat{sl_2})$. We then show how the generators of the
boundary quantum groups for the sine-Gordon model restricted to
the half-line by a general integrable boundary condition can be
constructed from these by using first order perturbation theory
\cite{Delius2}. These boundary quantum groups have been used to
determine $sl_{2}$ reflection matrices for arbitrary spin
\cite{Delius3}, to rederive the sine-Gordon soliton reflection
matrix \cite{Delius2} and to study dynamic boundaries
\cite{Baseilhac:2002kf}. Section~\ref{aT} generalizes the
calculation to affine Toda theories on the half-line
\cite{Delius2}, where it has been used to derive the reflection
matrices for the vector solitons in $a_{n}^{(1)}$ \cite{Delius2},
$d_{n}^{(1)}$ \cite{Delius4}, $c_{n}^{(1)}$ and $a_{2n-1}^{(2)}$
affine Toda theories. In section~\ref{last} we show that the half
line conserved charges described in sections~\ref{sG} and
\ref{aT}, where they were found using first order boundary
perturbation theory, are in fact exact \cite{Delius2}.

\section{The Yang-Baxter equation and quantum groups}
\label{YBE}
\EPSFIGURE{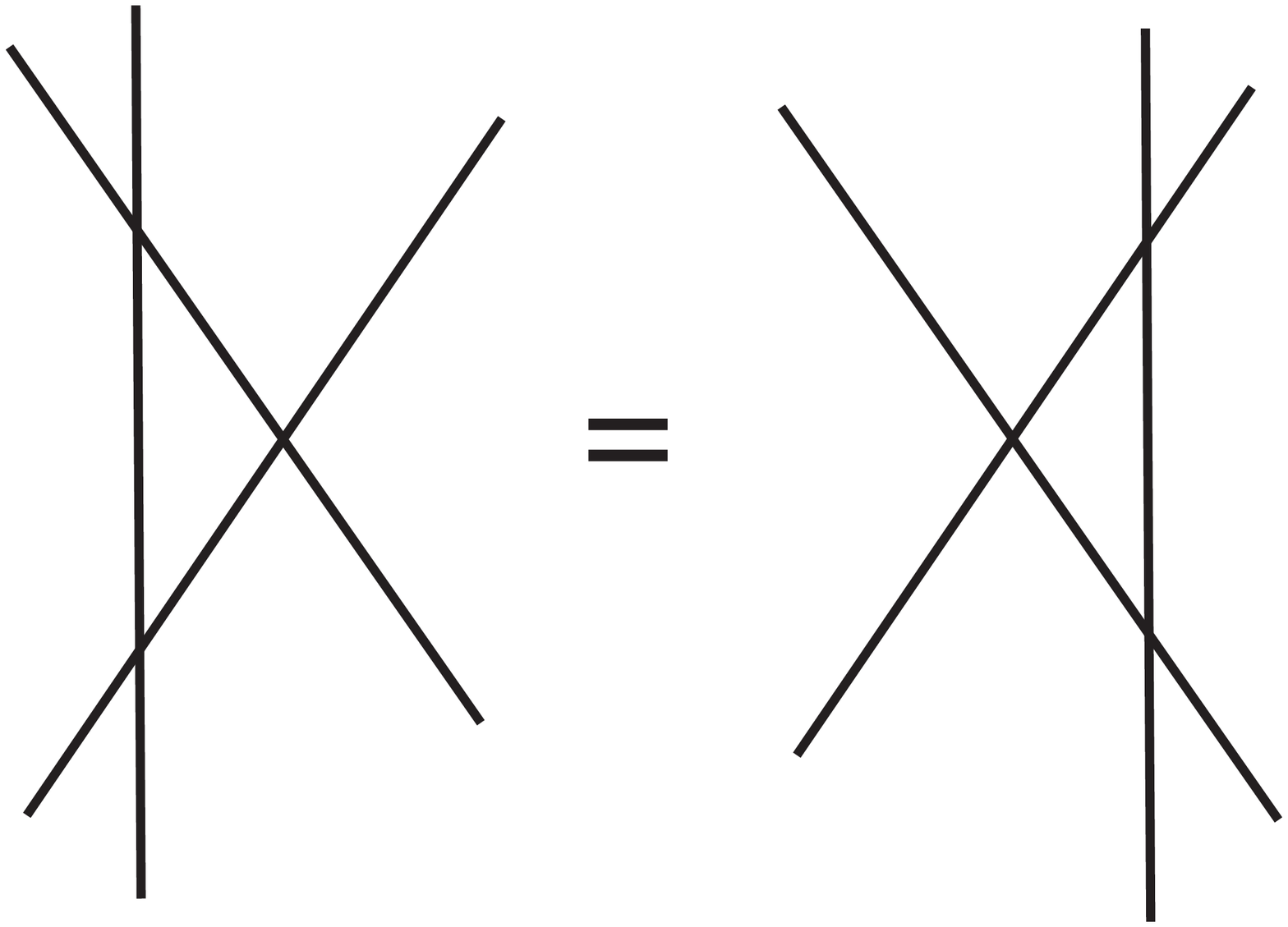, width= 5cm, height= 2.5cm}
{Yang-Baxter equation.\label{YBEfig}}
In this section we describe the Yang-Baxter equation using a
particle interpretation to aid clarity. Figure \ref{YBEfig} shows
two three-particle scattering processes in two dimensions. The two
scattering processes differ only in the order in which the
individual two-particle scatterings take place. In an integrable
theory the scattering amplitude will be independent of this order
and this is expressed by the equality in the figure. Each of the
lines in the figure stands for a vector space spanned by all the
quantum particle states in the same multiplet. Each vector space
carries a representation of the symmetry group describing a
particle multiplet. Let us denote the three vector spaces by
$V_{\theta_1}^{\mu}, V_{\theta_2}^{\nu}, V_{\theta_3}^{\lambda}$.
Here $\mu$, $\nu$ and $\lambda$ label the different multiplets and
the $\theta_i$ give the rapidities of the particles. Two or more
adjacent lines, for example those representing the spaces
$V^{\mu}_{\theta_1}$ and $V^{\nu}_{\theta_2}$, are tensor products
of those spaces, $V_{\theta_1}^{\mu}{\otimes}V_{\theta_2}^{\nu}$.

Crossing lines in the diagram represent scattering of two
particles and can be described by a mapping from the vector space
representing the initial two-particle state to the vector space
representing the final two-particle state,
\begin{equation}
\label{Rmatrix} \check{R}^{\mu \nu}(\theta_1-\theta_2) :
V^{\mu}_{\theta_1} \otimes V^{\nu}_{\theta_2}
    \to  V^{\nu}_{\theta_2} \otimes V^{\mu}_{\theta_1}.
\end{equation}
Because of Lorentz invariance the matrix $\check{R}^{\mu \nu}$
depends only on the difference $\theta_1-\theta_2$ of the spectral
parameters. This matrix, describing the scattering of two
particles, is usually referred to as the two-particle S-matrix.
The equation in figure \ref{YBEfig} becomes
\begin{multline}
\label{YBEq}
  \left(\check{R}^{\mu\nu}(\theta_{1}-\theta_{2})\otimes 1\right)
  \left(1\otimes\check{R}^{\mu\lambda}(\theta_{1}-\theta_{3})\right)
  \left(\check{R}^{\nu\lambda}(\theta_{2}-\theta_{3})\otimes 1\right)
  \\
   =
  \left(1\otimes\check{R}^{\nu\lambda}(\theta_{2}-\theta_{3})\right)
  \left( \check{R}^{\mu\lambda}(\theta_{1}-\theta_{3})\otimes 1\right)
  \left(1\otimes \check{R}^{\mu\nu}(\theta_{1}-\theta_{2})\right).
\end{multline}
This is the Yang-Baxter equation. We used the symbol $\check{R}$
to denote the scattering matrix because a solution of the
Yang-Baxter equation is conventionally referred to as an
$R$-matrix. Further details on $R$-matrices and the Yang-Baxter
equation can be found throughout the literature (for example
\cite{Baxter, Zamolodchikov2, Zamolodchikov1}).

The Yang-Baxter equation \eqref{YBEq} is non-linear, and so it is
difficult to solve directly. However in the physical context of
particle scattering in an integrable theory described above it is
possible that the scattering matrix is uniquely determined by the
symmetry of the theory. It is then enough to solve the symmetry
requirement in order to find a solution of the Yang-Baxter
equation. This is of great advantage because the symmetry
requirement corresponds to a simple linear equation. The quantum
groups generated by non-local conserved charges in certain
integrable quantum field theories describe such symmetries.
Examples are the Yangians $Y(\mathfrak{g})$ in the principal
chiral model
 and the quantum affine algebras $U_q(\hat{\mathfrak{g}})$ in affine Toda
field theory.

Let us denote the quantum group by $\mathcal{A}$.  The vector
spaces $V^{\mu}_{\theta}$ are representations of $\mathcal{A}$,
also called  $\mathcal{A}$-modules. The tensor product spaces
$V^{\mu}_{\theta_1}{\otimes}V^{\nu}_{\theta_2}$ formed by the
two-particle states should also be $\mathcal{A}$-modules.
Therefore $\mathcal{A}$ is required to be a bialgebra for which
there is defined a coproduct
\begin{equation}
\label{coproduct}
\triangle : \mathcal{A} \to \mathcal{A} \otimes \mathcal{A} \ .
\end{equation}
The coproduct encodes how a symmetry charge acts on multiparticle
states and in the case of non-local charges this is usually not
cocommutative.

Symmetry requires that the scattering matrix  \eqref{Rmatrix} is
 an $\mathcal{A}$-module homomorphism, also called an
intertwiner. Thus it has the property that for all
$Q{\in}\mathcal{A}$
\begin{equation}
\label{intertwiner} Q^{\nu \mu} \check{R}^{\mu \nu} =
\check{R}^{\mu \nu}Q^{\mu \nu}
\end{equation}
where $Q^{\mu \nu}$ denotes the action of $Q$ on
$V^{\mu}_{\theta_1}{\otimes}V^{\nu}_{\theta_2}$. We want the
quantum group symmetry to be strong enough to determine the
scattering matrix uniquely. Thus there should be only a unique
solution of the intertwining equation \eqref{intertwiner}. This is
ensured by Schur's lemma if the tensor product module
$V^{\mu}_{\theta_1}{\otimes}V^{\nu}_{\theta_2}$ is irreducible for
generic values of the spectral parameters. This is know to be the
case for Yangians $Y(\mathfrak{g})$ and quantum affine algebras
$U_q(\hat{\mathfrak{g}})$.

As stated above, in integrable quantum field theories the
scattering matrix $\check{R}^{\mu \nu}$ satisfies the Yang-Baxter
equation because the ordering of the two-particle scattering
processes does not matter. We can now give an independent
mathematical reason for why $\check{R}^{\mu \nu}$ satisfies the
Yang-Baxter equation. The tensor products
$V^{\mu}_{\theta_1}{\otimes}V^{\nu}_{\theta_2}{\otimes}V^{\lambda}_{\theta_3}$
are irreducible representations of $\mathcal{A}$, for generic
values of the spectral parameters. Thus the two sides of the
Yang-Baxter equation represent two possible maps from one
irreducible representation to another. Therefore, by Schur's
Lemma, they must be proportional to each other. The constant of
proportionality can be shown to be equal to $1$ by consideration
of the determinants of the two sides of \eqref{YBEq} and the
behaviour of the constant in the classical limit \cite{Jimbo}.

\section{The reflection equation and boundary quantum groups}
\label{bYBE}
\EPSFIGURE{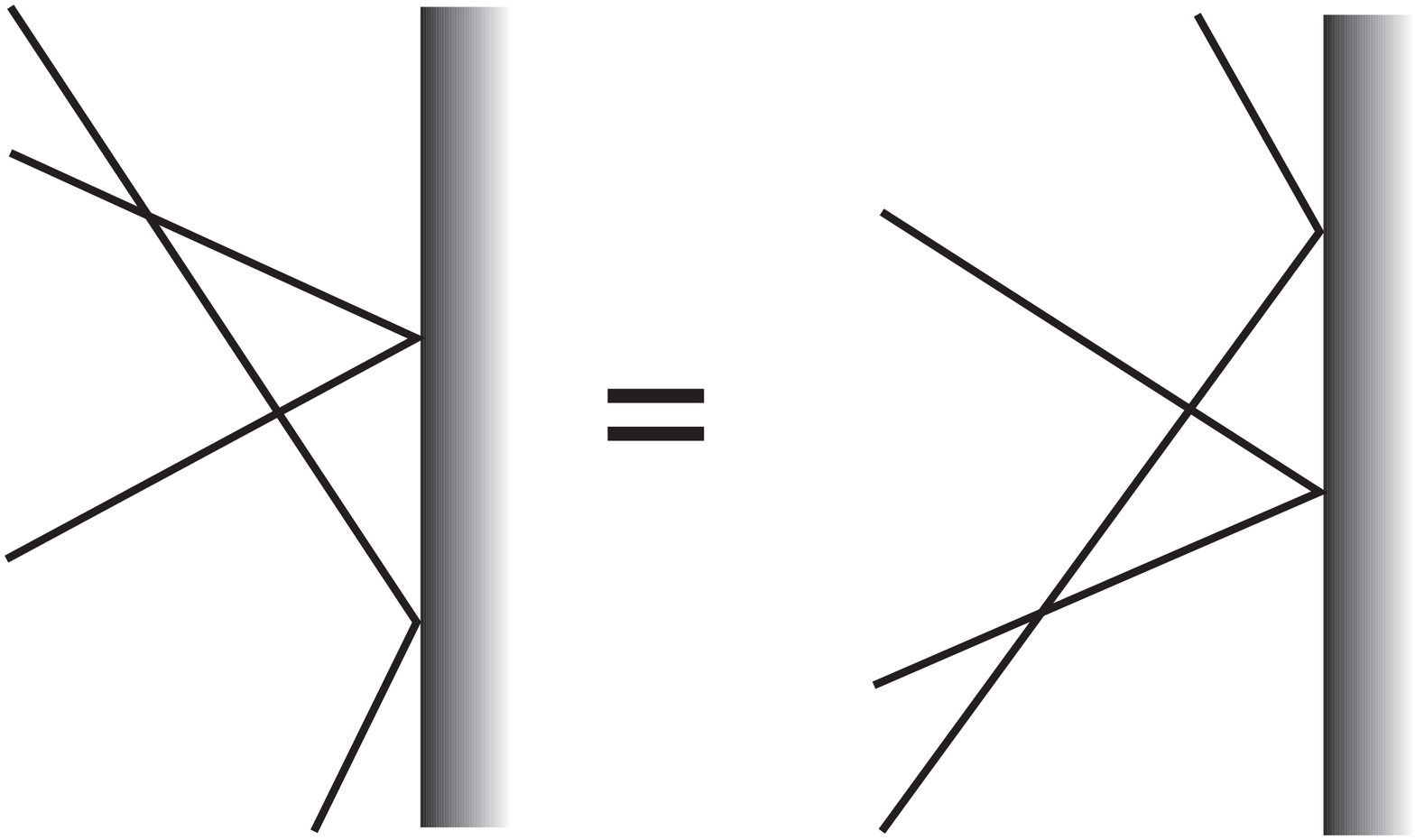, width=5cm, height=2.5cm}
{Reflection equation \label{bYBEfig}}
Figure \ref{bYBEfig} shows two different ways in which a
two-particle reflection process can factorize into single-particle
reflection and two-particle scattering processes. In integrable
quantum field theories the corresponding amplitudes must be equal.
The incoming lines represent vector spaces $V^{\mu}_{\theta_1},
V^{\nu}_{\theta_2}$. As in the Yang-Baxter equation the spaces
carry a representation of the symmetry algebra describing a
particle multiplet, labelled by the Greek letters. The spaces also
carry a spectral parameter $\theta_i$, giving the rapidity of the
corresponding particle. After reflection from the boundary the
particles will have reversed their rapidities and thus the
outgoing lines represent vector spaces
$V^{\bar{\mu}}_{-\theta_1}$, $V^{\bar{\nu}}_{-\theta_2}$. The
boundary also carries a vector space, $W^{\lambda}$, to describe
properties such as boundary spin or boundary bound states. Note
that this space does not have a rapidity parameter as the boundary
is assumed to be stationary. As in the Yang-Baxter equation
adjacent lines, including the boundary, represent tensor products
of the relevant vector spaces. The point of contact of the
boundary with and incoming and an outgoing line indicates
reflection of a particle from the boundary, this process can be
described with a mapping similar to \eqref{Rmatrix},
\begin{equation}
\label{Kmatrix}
  K^{\mu \nu}(\theta) : V^{\mu}_{\theta} \otimes W^{\lambda}
  \to V^{\bar{\mu}}_{-\theta} \otimes W^{\lambda}.
\end{equation}
$K^{\mu \nu}$ are the reflection matrices or K-matrices.
Intersection of two lines, incoming or outgoing, represents
scattering of the particles and, as in the previous section, is
represented by $\check{R}$. The equation in figure \ref{bYBEfig}
then reads
\begin{multline}
\label{bYBEq}
  \left(\check{R}^{\bar{\nu} \bar{\mu}}(\theta_1 -
  \theta_2)\otimes 1\right)
  \left( I \otimes K^{\mu \lambda}(\theta_{1}) \right)
  \left(\check{R}^{\mu \bar{\nu}}(\theta_1 + \theta_2 )\otimes 1\right)
  \left( I \otimes K^{\nu \lambda}(\theta_{2} ) \right)
  \\=
  \left( I \otimes K^{\nu \lambda}(\theta_{2} ) \right)
  \left(\check{R}^{\nu \bar{\mu}}(\theta_1 + \theta_2 )\otimes 1\right)
  \left( I \otimes K^{\mu \lambda}(\theta_1 ) \right)
  \left(\check{R}^{\mu \nu}(\theta_1 -\theta_2 )\otimes 1\right).
\end{multline}
This is the reflection equation. Further details on the reflection
equation can be found throughout the literature (for example
\cite{Cherednik, Ghoshal, Sklyanin}).

The reflection equation \eqref{bYBEq} is non-linear. As so many
other similarities hold between this and the Yang-Baxter equation
it might be expected that a method analogous to the intertwiners,
as described in the previous section, may exist to find
$K$-matrices by the solution of a linear equation. To this end we
are motivated to find boundary quantum groups, $\mathcal{B}$,
analogous to $\mathcal{A}$, the quantum group of the bulk theory
\cite{Sklyanin}. $\mathcal{B}$ is the symmetry algebra of the
boundary theory. It is generated by the non-local charges that are
still conserved after the introduction of the boundary. The vector
space associated with the boundary $W^{\lambda}$ is a
representation of the algebra $\mathcal{B}$, called a
$\mathcal{B}$-module. The tensor product
$V^{\mu}_{\theta}{\otimes}W^{\lambda}$, describing a particle and
the boundary, must also be a $\mathcal{B}$-module. We know that
the space $V^{\mu}_{\theta}$ is an $\mathcal{A}$-module so
$\mathcal{B}$ is required to be an $\mathcal{A}$ coideal, for
which there exists a coproduct,
\begin{equation}
\label{coprodb}
\triangle : \mathcal{B} \to \mathcal{A} \otimes \mathcal{B} \ .
\end{equation}
This coproduct describes the action of the non-local charges on
states containing particles and the boundary. This observation
explains why the symmetry algebra of an integrable quantum field
theory with a boundary does not have to be a bialgebra: the
particles away from the boundary still transform in
representations of the symmetry of the theory on the whole line.

Just as the $R$-matrix was an $\mathcal{A}$-module homomorphism it
is now clear  that the symmetry requires that the $K$-matrix is a
$\mathcal{B}$-module homomorphism, or intertwiner. So for all
$\hat{Q}{\in}\mathcal{B}$ the $K$-matrix has to satisfy
\begin{equation}
\label{bintertwiner} \hat{Q}^{\bar{\mu} \lambda} K^{\mu \lambda} =
K^{\mu \lambda} \hat{Q}^{\mu \lambda},
\end{equation}
where $\hat{Q}^{\mu \lambda}$ denotes the action of $\hat{Q}$ on
$V^{\mu}_{\theta} \otimes W^{\lambda}$.

If
$V^{\mu}_{\theta_1}{\otimes}V^{\nu}_{\theta_2}{\otimes}W^{\lambda}$
are irreducible representations, for generic $\theta_i$, then by
Schur's Lemma a solution of the linear intertwining equation
\eqref{bintertwiner} is also a solution of the reflection equation
\eqref{bYBEq}. The argument is exactly analogous to the one
presented in the previous section to show that a solution of eq.
\eqref{intertwiner} is a solution of the Yang-Baxter equation.

The problem of finding reflection matrices has now been reduced to
finding the boundary quantum group, and the solution of the linear
equation \eqref{bintertwiner}. The following properties have to
hold in order for $\mathcal{B}$ to lead us to solutions of the
reflection equation:
\begin{itemize}
\item $\mathcal{B}$ must be an $\mathcal{A}$ coideal, \item
Intertwiners $ K^{\mu \lambda}(\theta):
      V^{\mu}_{\theta}{\otimes}W^{\lambda} \to
      V^{\bar{\mu}}_{-\theta}{\otimes} W^{\lambda}$ must exist,
\item the tensor product representations
$V^{\mu}_{\theta_1}{\otimes}V^{\nu}_{\theta_2}{\otimes}W^{\lambda}$
      must be irreducible for generic values of the
      spectral parameters $\theta_i$.
\end{itemize}
We will now go on to find such boundary quantum groups by studying
the symmetries in concrete integrable quantum field theories on
the half-line.

\section{Sine-Gordon field on the half line}
\label{sG}

We seek to find the boundary quantum group for the sine-Gordon
field restricted to the half line by an integrable boundary
condition. Initially, however, we consider the whole line theory.
It is useful to view the sine-Gordon model as a perturbation of
the free bosonic field theory, with action \cite{Zamolodchikov1,
Bernard} \footnote{We use the conventions of \cite{Bernard}. The
contribution to these proceedings by Bajnok, Palla, and Tak\'{a}cs
on the boundary sine-Gordon model \cite{Baj} uses more standard
conventions. The relations are that
$\hat{\beta}=\beta/\sqrt{4\pi}$, $\phi=\sqrt{4\pi}\Phi$, and
$\lambda=2\pi m^2/\beta^2$.}
\begin{equation}
\label{sGaction}
S = \frac{-1}{8 \pi} \iint \left( ( \partial_x \phi)^2
                          + (\partial_t \phi)^2 \right) \ dx\ dt
  - \Phi^{\text{pert}}_{\text{bulk}} \ ,
\end{equation}
where
\begin{equation}
\label{sGpert}
\Phi^{\text{pert}}_{\text{bulk}} = \frac{\lambda}{4 \pi} \iint
\left( e^{i
\hat{\beta} \phi (x,t)} + e^{-i \hat{\beta} \phi(x,t)}
\right) \ dx \ dt \ .
\end{equation}
The model has non-local conserved charges $Q_{\pm}$,
$\bar{Q}_{\pm}$, their expressions can be found in \cite{Bernard}.
These charges together with the Lorentz boost $D$ and the
topological charge
\begin{equation}
\label{sGtop} T = \frac{ \hat{\beta} }{2 \pi} \int \partial_x \phi
\ dx
\end{equation}
generate the quantum affine algebra $U_q(\widehat{sl_2})$ with
zero center.

%
The action of the non-local conserved charges on two-soliton
states are given by the coproduct
\begin{align}
\nonumber
\triangle & \left( Q_{\pm} \right) =
Q_{\pm} \otimes I + q^{\pm T} \otimes Q_{\pm} \ , \\
\label{coprod1}
\triangle & \left( \bar{Q}_{\pm} \right) =
\bar{Q}_{\pm} \otimes I + q^{\mp T} \otimes \bar{Q}_{\pm} \ , \\
\nonumber \triangle & \left( T \right) = T \otimes I + I \otimes
T,
\end{align}
where
\begin{equation}
\label{q1} q = e^{2i \pi ( 1- \hat{\beta}^{2} )/ \hat{\beta}^{2}}.
\end{equation}

We will treat the sine-Gordon model on the half-line with general
integrable boundary conditions as a perturbation of the free field
theory on the half line with Neumann boundary condition
$\partial_x{\tilde{\phi}}|_{x=0}=0$. Note that to avoid confusion
we decorate the fields in the theory on the half-line with a
tilde. The Neumann boundary condition selects among all possible
solutions on the whole line exactly those that are invariant under
parity $\mathcal{P}:x\mapsto -x$. We find that the easiest way to
do quantum calculations in the theory with Neumann boundary
condition is to view it as the parity invariant subsector of the
theory on the whole line.

The perturbing term $\Phi^{\text{pert}}_{\text{bulk}}$ in
\eqref{sGpert} is invariant under parity transform, $\mathcal{P}$,
and so we can also use it as the perturbing term to obtain the
sine-Gordon model on the half-line with Neumann boundary
condition.

The whole line conserved charges transform under parity as
\begin{align}\label{Ptrans}
  \mathcal{P}: \ T &\mapsto -T,&
  \mathcal{P}: \ Q_{\pm} &\mapsto \bar{Q}_{\mp},&
  \mathcal{P}: \ \bar{Q}_{\pm} &\mapsto Q_{\mp}.
\end{align}
The parity invariant combinations
$\tilde{Q}_{\pm}=Q_{\pm}+\bar{Q}_{\mp}$ are well defined in the
half-line theory and are conserved \cite{Delius2}.

We now replace the Neumann condition with the more general
integrable boundary condition \cite{Ghoshal}
\begin{equation}
\label{GZBC}
\partial_x \tilde{\phi}(x,t)|_{x=0} =
i \hat{\beta} \lambda_b \left( \epsilon_{-} e^{i \hat{\beta}
\tilde{\phi}(0,t) / 2} - \epsilon_{+} e^{-i \hat{\beta}
\tilde{\phi}(0,t) / 2} \right),
\end{equation}
where $\epsilon_+$ and $\epsilon_-$ are two arbitrary boundary
parameters \footnote{These are related to the boundary parameters
$M_0$ and $\varphi_0$ used in \cite{Baj} by
$\exp(i\beta\varphi_0)=\epsilon_+/\epsilon_-$ and
$M_0^2=(\lambda_b/\pi)^2\epsilon_+\epsilon_-$.}.
We treat this as a boundary perturbation on the Neumann half line problem
discussed above, with action $S_{\epsilon}$ given by
\begin{equation}
\label{GZBCaction}
S_{\epsilon} = S_{0} +
\Phi^{\text{pert}}_{\text{bound}} \ ,
\end{equation}
where $S_{0}$ is the action of the Neumann boundary theory
and
\begin{equation}
\label{boundpert}
\Phi^{\text{pert}}_{\text{bound}} = \frac{\lambda_b}{2 \pi} \int
\left(
\epsilon_{-} e^{i \hat{\beta} \tilde{\phi} (0,t) /2} +
\epsilon_{+} e^{-i \hat{\beta} \tilde{\phi} (0,t) /2}
\right) \ dt \ .
\end{equation}
We can then use first order boundary perturbation theory \cite{Ghoshal,
Penati} to find the half
line conserved charges \cite{Delius2}
\begin{equation}
\label{GZBCcharges}
\hat{Q}_{\pm} = Q_{\pm} + \bar{Q}_{\mp} + \hat{\epsilon}_{\pm} q^{\pm
T}\ ,
\end{equation}
where
\begin{equation}
\label{epsilon}
\hat{\epsilon}_{\pm} = \frac{ \lambda_{b} \epsilon_{\pm} \hat{\beta}^{2}}
                            {2 \pi (1- \hat{\beta}^{2})} \ .
\end{equation}
These expressions were first conjectured in \cite{Mezincescu}. We denote
the symmetry algebra generated by these charges by
$\mathcal{B}^{\epsilon}_{q}(\widehat{sl_2})$.
$\mathcal{B}^{\epsilon}_{q}(\widehat{sl_2})$ is a coideal subalgebra of
$U_q(\widehat{sl_2})$ because
\begin{equation}
\label{subalg}
\triangle \left( \hat{Q}_{\pm} \right) =
\hat{Q}_{\pm} \otimes I + q^{\pm T} \otimes \left( \hat{Q}_{\pm} -
\hat{\epsilon}_{\pm} I \right) \in U_{q}(\widehat{sl_2}) \otimes
\mathcal{B}^{\epsilon}_q (\widehat{sl_2}) \ .
\end{equation}

This boundary quantum group can now be used to calculate
reflection matrices by solving the intertwining relation
\eqref{bintertwiner}. Choosing the representation $W^\lambda$ to
be the trivial representation (corresponding to the boundary
ground state) and the representation $V^\mu$ to be the spin $1/2$
representation (in which the sine-Gordon solitons are known to
transform) one rederives the known sine-Gordon soliton reflection
matrix of Ghoshal and Zamolodchikov \cite{Ghoshal}. The relations
between our parameters $\hat{\epsilon}_\pm$ and the parameters
$\eta$ and $\Theta$ in \cite{Ghoshal} are
$\hat{\epsilon}_\pm=f(\lambda,\hat{\beta})\,\cos(\eta\pm i\Theta)$
where $f(\lambda,\hat{\beta})$ is a complicated function of
$\lambda$ and $\hat{\beta}$. By taking the representation $V^\mu$
to be higher spin representations of $U_q(\widehat{sl_2})$ one
obtains new $sl_2$ $K$-matrices \cite{Delius3}.

\section{Affine-Toda fields on the half line}
\label{aT}

We will now follow a similar procedure to find boundary quantum
groups $\mathcal{B}^{\epsilon}_{q}(\hat{\mathfrak{g}})$ for every
affine Lie algebra $\hat{\mathfrak{g}}$ by deriving the non-local
symmetry charges in affine Toda field theories.

For every affine Lie algebra $\hat{\mathfrak{g}}$ of rank $n$
there is an affine Toda field theory describing a $n$-component
bosonic field in 1+1 dimensions with action \cite{Mikhailov}
\begin{equation}
\label{aTaction}
S = \frac{-1}{8 \pi} \iint
\left( ( \partial_x \phi )^{2} + ( \partial_t \phi
)^{2}
\right) \ dx \ dt -\Phi^{\text{pert}}_{\text{bulk}} \ ,
\end{equation}
where
\begin{equation}
\label{aTpert} \Phi^{\text{pert}}_{\text{bulk}} = \frac{\lambda}{4
\pi} \iint \left( \sum^{n}_{j=0} e^{-i \beta \alpha_j \cdot \phi /
| \alpha_j |^{2} } \right) \ dx \ dt.
\end{equation}
The $\alpha_j$ are the simple roots of $\hat{\mathfrak{g}}$
projected onto the root space of $\mathfrak{g}$, the finite
dimensional Lie algebra. The theory has non-local conserved
charges $Q_j$ and $\bar{Q}_j$ for $j=0,1,\dots,n$ \cite{Bernard}.
These, along with the Lorentz boost $D$ and the topological
charges
\begin{equation}
\label{aTtop} T_j = \frac{ \beta }{2 \pi} \int \alpha_j \cdot
\partial_x \phi \ dx,
\end{equation}
generate the quantum affine algebra $U_q(\hat{\mathfrak{g}})$ with
zero centre \cite{Felder}.
The non-local charges act on two-soliton states through the
coproduct
\begin{align}
\nonumber
\triangle & \left( Q_j \right) =
Q_j \otimes I + q^{T_j} \otimes Q_j \ , \\
\label{coprod2}
\triangle & \left( \bar{Q}_j \right) =
\bar{Q}_j \otimes I + q^{T_j} \otimes \bar{Q}_j \ , \\
\nonumber
\triangle & \left( T_j \right) =
T_j \otimes I + I \otimes T_j \ .
\end{align}
For $\hat{\mathfrak{g}}=\widehat{sl}_2$ the affine Toda theory
becomes the sine-Gordon model and the conserved non-local charges
can be equated to those in section \ref{sG} as $Q_{0}=Q_{+}$,
$Q_{1}=Q_{-}$, $\bar{Q}_{0}=\bar{Q}_{-}$,
$\bar{Q}_{1}=\bar{Q}_{+}$ and $T_{0}=-T_{1}=T$.

We are interested in restricting the field, for simply laced
$\hat{\mathfrak{g}}$, to the half line with the
boundary condition
\begin{equation}
\label{aTbc}
\partial_x \tilde{\phi}(x,t)|_{x=0} =
- i \beta \lambda_b \sum_{j=0}^{n} \epsilon_{j} \alpha_{j} e^{- i
\beta \alpha_j \cdot \tilde{\phi}(0,t) / 2 } \ ,
\end{equation}
where the $\epsilon_j$ are $n+1$ free boundary parameters. It is
known that these boundary conditions preserve the integrability of
the theory only if all $|\epsilon_{j}|=1$ or if all
$\epsilon_{j}=0$ \cite{Bowcock}.

As for the sine-Gordon field we define the Neumann half line
theory using parity invariance. We then add a boundary
perturbation giving the action,
\begin{equation}
\label{aTboundaction}
S_{\epsilon} = S_0 +
\Phi^{\text{pert}}_{\text{bound}} \ ,
\end{equation}
where $S_{0}$ is the action of the Neumann theory and
\begin{equation}
\label{aTboundpert}
\Phi^{\text{pert}}_{\text{bound}} = \frac{ \lambda_b}{2 \pi} \int
\left(
\sum_{j=0}^{n} \epsilon_j e^{- i \beta
\alpha_j \cdot \tilde{\phi} (0,t) / 2} \right) \ dt \ .
\end{equation}
Using first order boundary perturbation theory the non-local conserved
charges have been found to be \cite{Delius2}
\begin{equation}
\label{aTcharges} \hat{Q}_j = Q_j + \bar{Q}_j + \hat{\epsilon}_j\
q_{j}^{T_j} \ ,
\end{equation}
where
\begin{equation}
\label{aTepsilon} \hat{\epsilon}_j = \frac{ \lambda_{b}\,
\hat{\beta}^{2}}
                        { 2 \pi (1- \hat{\beta}^{2})}\,\epsilon_j.
\end{equation}
While we have performed the calculation only in the case of simply
laced $\hat{\mathfrak{g}}$, we believe the expressions
\eqref{aTcharges} to be correct in all cases.

 We denote the symmetry algebra generated by the charges
$\hat{Q}_j$, $j=0,1,\dots,n$ by
$\mathcal{B}^{\epsilon}_{q}(\hat{\mathfrak{g}}){\subset}U_{q}(\hat{\mathfrak{g}})$.
In \cite{Delius4} we introduced the name `quantum affine
reflection algebras' for these algebras to express that they are
subalgebras of quantum affine algebras which at the same time are
reflection equation algebras in the sense of Sklyanin
\cite{Sklyanin}. The  algebra
$\mathcal{B}^{\epsilon}_{q}(\hat{\mathfrak{g}})$ is a coideal
subalgebra of $U_{q}(\hat{\mathfrak{g}})$ because
\begin{equation}
\label{cidsubalg} \triangle \left( \hat{Q}_{j} \right) =
\hat{Q}_{j} \otimes I + q^{T_j} \otimes \left( \hat{Q}_{j} -
\hat{\epsilon}_j I \right) \in U_q(\hat{\mathfrak{g}}) \otimes
\mathcal{B}^{\epsilon}_{q}(\hat{\mathfrak{g}}),
\end{equation}
as can easily be checked using \eqref{coprod2}.

To first order in boundary perturbation theory the charges
$\hat{Q}_j$ are conserved for arbitrary values of the boundary
parameters $\epsilon_j$. However we found that the intertwining
relation \eqref{bintertwiner} does not always have a solution. In
fact we found that for simply laced $\hat{\mathfrak{g}}$ the
boundary parameters $\hat{\epsilon}_j$ are fixed up to sign by the
requirement that a reflection matrix for the vector representation
should exist. This is the quantum generalization of the result of
\cite{Bowcock} which showed that the boundary condition preserves
classical integrability only if the boundary parameters
$\epsilon_j$ are fixed up to sign. For non-simply laced
$\hat{\mathfrak{g}}$ we find that some of the parameters
$\hat{\epsilon}_j$ can be free and this again parallels the
classical results of \cite{Bowcock}.

The boundary quantum groups
$\mathcal{B}^{\epsilon}_{q}(\hat{\mathfrak{g}})$ have been
successfully used to rederive the vector representation reflection
matrices for $a_{n}^{(1)}$ affine Toda theories \cite{Delius2} and
to find previously unknown reflection matrices for $d_{n}^{(1)}$
\cite{Delius4}, $c_{n}^{(1)}$ and $a_{2n-1}^{(2)}$ affine Toda
theories.

\section{Non-perturbative derivation of non-local conserved charges}
\label{last}

In sections \ref{sG} and \ref{aT} we have described how the
boundary quantum group charges can be found using first order
boundary perturbation theory. One might be worried that at higher
order perturbation theory additional terms could appear in the
expression for the half line conserved charges. The fact that the
charges have already been used successfully to derive new
$K$-matrices for the vector representation does not rule out the
presence of additional terms because these terms could vanish in
the vector representations. We can, however, derive the expression
for the charges also in a non-perturbative way by using the vector
soliton reflection matrix. This derivation, which we will present
briefly in this section, shows that the expressions
\eqref{aTcharges} for the symmetry charges are exact and only the
relation \eqref{aTepsilon} between the bare and the renormalized
boundary parameters may receive higher order corrections
\cite{Delius2}.

Using the vector soliton reflection matrix $K^{\mu}(\theta)$ we
construct a matrix
\begin{equation}
\label{Bmatrix} B^{\mu}_{\theta} = \bar{L}_{\theta}^{\bar{\mu}} (
K^{\mu} (\theta) \otimes 1) L_{\theta}^{\mu} \in \text{Hom}
(V^{\mu}_{\theta}, V^{\bar{\mu}}_{-\theta}) \otimes
U_{q}(\hat{\mathfrak{g}}) \ ,
\end{equation}
with
\begin{align}
\label{L1} L^{\mu}_{\theta} &= ( \pi^{\mu}_{\theta} \otimes id) (
\mathcal{R}(x) ) \in \text{End} (V^{\mu}_{\theta} )
\otimes U_{q}(\hat{\mathfrak{g}}) \ , \\
\label{L2} \bar{L}^{\bar{\mu}}_{\theta} &= (
\pi^{\bar{\mu}}_{-\theta} \otimes id) (\mathcal{R}^{op}(x)) \in
\text{End}(V^{\bar{\mu}}_{-\theta} ) \otimes
U_{q}(\hat{\mathfrak{g}}) \ ,
\end{align}
where $\pi^{\nu}_{\theta}$ denotes the representation
$\pi^{\nu}_{\theta}:U_{q}(\hat{\mathfrak{g}})\to
\text{End}(V^{\nu}_{\theta})$,
$\mathcal{R}(x)=(\Psi_x{\otimes}id)(\mathcal{R})$, $\Psi_x$ is a
homomorphism defined by $\Psi_x(Q_j)=xQ_j$,
$\Psi_x(\bar{Q}_j)=x^{-1}\bar{Q}_j$, $\Psi_x(T_j)=T_j$,
$\mathcal{R}$ is the universal $R$-matrix \cite{Khoroshkin} and
$\mathcal{R}^{op}$ is the opposite universal $R$-matrix found by
swapping the tensor factors in $\mathcal{R}$. It can now be shown
that any $K$-matrix that satisfies the reflection equation
\eqref{bYBEq} commutes with all elements of $B^{\mu}_{\theta}$ for
any $\theta$ \cite{Delius2}. Thus if we expand $B^{\mu}_{\theta}$
in terms of the spectral parameter, then every term in the
expansion will commute with any $K$-matrix. Thus every term in the
expansion is a symmetry charge. In the cases of
$\hat{\mathfrak{g}}=a_n^{(1)}$ and $\hat{\mathfrak{g}}=d_n^{(1)}$
we have checked explicitly that the first term in this expansion
reproduces exactly the charges \eqref{aTcharges} derived
previously.

A similar construction to the above has recently also been used in
\cite{Mol} to study other properties of the boundary quantum
groups.

\section{Summary and Outlook}
\label{conc}

We have presented a brief review of the Yang-Baxter equation
describing how the concept of quantum group arises and how
solutions of the Yang-Baxter equation can be found as intertwiners
of representations of that group. We also presented a review of
the reflection equation and introduced the concept of a boundary
quantum group as a coideal subalgebra of the quantum group. We
showed that solutions to the reflection equation can be found as
intertwiners of representations of the boundary quantum group. We
showed how the conserved non-local charges for affine Toda
theories on the half line can be found using first order
perturbation theory, and how these generate boundary quantum
groups. Finally we showed that these conserved charges are exact.

There is much work remaining in this area, reflection matrices for
the vector representations of $b_{n}^{(1)}$, $a_{2n}^{(2)}$,
$d_{n+1}^{(2)}$ and the exceptional algebras have yet to be found.
Boundary quantum groups corresponding to Dirichlet type boundary
conditions need to be studied \cite{Nep}, as do higher
representations of the presently known groups.

\end{document}